\documentclass[twoside]{article}
\usepackage{fleqn,espcrc2,epsfig}

\usepackage{graphicx}

\def\spa{\hspace*{-.65cm}}
\def\eqs#1#2{eqs.~(\ref{#1}--\ref{#2})}
\def\eg{\epsilon_\gamma}
\def\ez{\epsilon_Z}
\newcommand{\db}{\hspace{-0.2ex}\not\hspace{-0.7ex}D\hspace{0.1ex}}
\newcommand{\eq}[1]{eq.~(\ref{#1})}

\newcommand{\AmS}{{\protect\the\textfont2
  A\kern-.1667em\lower.5ex\hbox{M}\kern-.125emS}}

\title{Bounds on the Tau Magnetic Moments: Standard Model and Beyond}

\author{Gabriel A. Gonz\'alez-Sprinberg\address[IFFC]{Instituto de 
F\'{\i}sica,
Facultad de Ciencias, Universidad de la Rep\'ublica,\\ 
Igu\'a 4225, 11400 Montevideo,  Uruguay},
        Arcadi Santamaria\address[VAL]{
Departament de F\'{\i}sica Te\`orica, IFIC, 
CSIC-Universitat de Val\`encia\\ E-46100 Burjassot, Val\`encia, Spain
} and 
      Jorge Vidal\addressmark[VAL]\thanks{Work supported by 
CICYT-Grant AEN-99-0692, by DGESIC-Grant PB97-1261, by DGEUI of 
the Generalitat Valenciana under the Grant GV98-01-80 and 
by Agencia Espa\~nola de Cooperaci\'on Internacional.}}
       
\begin{document}

\begin{abstract}We obtain new bounds for the magnetic dipole 
moments of the tau lepton. These limits on the magnetic 
couplings of the tau 
to the electroweak gauge bosons ($\gamma , W , Z$)  are set in a 
model independent way using the most general effective 
Lagrangian with the $SU(2)_L 
\otimes U(1)_Y$ symmetry. Comparison with data from the most 
precise experiments at high energies  shows that the present
limits are 
more stringent than the previous published ones. For the anomalous
magnetic moment the bounds are, for the first time, within one order of 
magnitude of the standard model prediction.
 \vspace{1pc}
\end{abstract}

\maketitle

\section{INTRODUCTION}

The fermion dipole moments are some of the best 
measured quantities in physics. While the limits for the 
CP-violating ones, the electric dipole moments, are 
impressive\cite{caso}
( $10^{-26} - 10^{-25}$ e-cm for the electron and the neutron), the
agreement between the measurement of the anomalous magnetic dipole 
moments of the electron and of the muon and its theoretical prediction is 
such that provides the best determination of the fine 
structure constant.  Since the first prediction for the electron 
anomalous 
magnetic moment computed by Schwinger\cite{schw}, the magnetic
 properties of the 
leptons played a central role in testing symmetries and 
the quantum prediction of the theory. This could be done
 thanks to the
relativity long lifetime of the muon and the stability of 
the electron in
ordinary atoms.    For the heavy fermions the situation is 
very different. 
As short lived-particles, there is no possibility of such 
high precision 
experiments. Nevertheless, the magnetic dipole moments are,
 for
on-shell particles,  physical magnitudes that not only contain 
important information about the interactions, but may also give 
some insights on the
mechanism of mass generation. Due to the high mass 
of the tau lepton (and other heavy fermions) the measurement of the 
dipole moments is an interesting test for the standard model and 
possible new physics that may show up there. 
  As expressed  by M.Perl\cite{perl} in the section {\it Dreams and
 odd ideas
in tau research}: "... It would be very nice to measure $\mu_\tau$ 
with enough precision to check this [the Schwinger term 
$\alpha /2\pi$], as it was 
checked for the $e$ and the $\mu$
years ago.{\em At present such precision is a dream."}
 Previous  bounds on the tau magnetic moment
$a_\gamma$ are  indirect \cite{masso}, or include a partial 
analysis of the amplitudes in the considered processes\cite{mendez,swain}. 
In the first case they come 
from the observation that in general extensions of the standard model 
(SM) it is very difficult to generate a weak magnetic moment, $a_Z$,
 for a lepton without originating a  coupling of the Z boson
 to the lepton  of the same order of magnitude. This  weak magnetic
 coupling is strongly bounded by LEP1, therefore,
by assuming that the magnetic moment of the lepton ($a_\gamma$) has 
the same size, one obtains a rather strong bound on it. On the 
other hand, far from the resonance the $a_\gamma$ magnetic moment couplings 
of the tau is more important and should be taken 
into account in the theoretical and experimental analysis.
 Finally, the tau-W magnetic coupling was first considered in the W
decays
 into tau plus leptons\cite{rizzo} and also at low energies in
the tau decays into leptons and pions\cite{low}.
 
While these arguments are plausible, a complete analysis of the 
 data coming
from tau-lepton production at LEP1, SLD and LEP2, and data on W 
decays into tau
leptons from LEP2 and $p\bar{p}$ colliders, allows for a better understanding 
of the magnetic moment couplings of the tau to the photon, the Z and the W 
bosons. 
More technical details of what follows can be found in
\cite{npb}.

\subsection{Dipole moments in the standard model and beyond}
The anomalous magnetic moments of the tau lepton, as the rest of the leptons, 
receive the standard contribution from the Schwinger term. However, in the
case of the tau lepton, larger contributions from new physics are still 
allowed. At the electroweak scale those can be studied within the effective
Lagrangian approach. 
Magnetic dipole moments are usually defined with all particles in the magnetic
moment vertex on-shell because {\em these
are the gauge invariant quantities of the theory, as shown in the 
classical paper} \cite{leesanda}.

For the $a_\gamma$ tau magnetic dipole moment there
is no available experiment satisfying the above requirement and 
to determine it one has to resort to general and consistent (gauge invariant) 
parameterizations of all physics in which the magnetic moment could play
a significant role. The most general of those parameterizations is based
on the effective field theory language which will allow us to relate 
off-shell matrix elements with the physical on-shell  anomalous magnetic 
 moment $a_\gamma$.
Instead, the $a_Z$ can be determined on-shell: we have tau pair production 
at the Z-peak at various experiments and specific observables proportional
to the weak magnetic moment have been calculated\cite{nos} and
measured\cite{l3}.

The effective Lagrangian approach \cite{buch} we use is 
described in section 2 where 
we also fix the notation. Magnetic moment are generated by
non-renormalizable couplings, their 
effects grow with energy and, therefore, high energy 
experiments, though not as precise as experiments at 
lower energies, can still be relevant. This is the 
case for LEP1 and SLD measurements as
 compared to LEP2 ones. 
 Besides, while the former are only sensitive 
to the Z-magnetic couplings, LEP2 is sensitive to both
 Z and photon magnetic moments ($a_Z$ and $a_\gamma$). 
Far from the resonance the statistics 
rapidly decreases and the precision is not as good. 
In general, lower energy experiments  do not provide stringent
 bounds: the suppression factor
$(E_{low}/m_Z)^2$ has to be compensated by higher precision in the 
experiment (see for instance the
bounds obtained from tau decays \cite{low}).

Observables are studied in section 3 where tau  production
 is considered in
  universality test and  transverse polarization 
at LEP1-SLD, total production rates 
at LEP2,
 and lepton universality
in $W$ decays in LEP2 and hadron colliders.

What are the observables related to  the magnetic moments? There is 
no symmetry guide as, for example, in CP-violation by 
electric dipole moments, where specific CP-odd 
observables can be 
constructed. However, magnetic couplings flip chirality,
 so  in order to select the
physical effects of the magnetic moments one should take 
these properties into account. 
In the standard model the only source of chirality flips are 
fermion masses, much lower
than the electroweak scale. This means that any contribution 
 to inclusive observables 
 are either suppressed
by the fermion masses or need two operator insertions and then, they come 
as the electroweak magnetic moments squared.
 Furthermore, any new physics terms,
 not only that 
related to the magnetic moments will appear in these  observables. 
It is therefore
convenient to also include observables that are exactly zero when 
chirality is conserved: these
will be only sensitive to fermion masses and/or magnetic moments and 
will depend linearly on the couplings.
Up to now,  observables 
of this kind have been studied at LEP1 but not at LEP2.

There are only two gauge invariant dimension six operators contributing
to magnetic moments in the effective Lagrangian. Therefore gauge invariance
imposes a relationship among the three weak and electromagnetic magnetic
moments.Therefore one can achieve 
some insight on tau magnetic moments by studying W decays into tau leptons.
There exist good bounds
on universality of W decays into leptons coming from LEP2, UA1, UA2, 
CDF and D0 that can be translated into limits on the magnetic moments.

In section 4 we present the combined analysis of the observables, compare
them to other bounds found in the literature and discuss 
the results. 

\section{NEW PHYSICS AND THE EFFECTIVE LAGRANGIAN}

The new physics phenomenology at low energy, {\it i.e.} at energies much lower
than the scale of this new physics,  can be
parametrized by an effective Lagrangian built with the standard model
particle spectrum and gauge symmetry, having as zero order term just 
the standard model
Lagrangian, plus higher dimension gauge invariant operators  
suppressed by the scale  of new physics $\Lambda$. The leading non-standard 
effects will come from the operators with the lowest dimension. Those are 
dimension six operators and among them there are only two operators
that contribute to the tau magnetic moments:
\begin{equation}
\label{eq:ob}
{\cal O}_B = \frac{1}{\Lambda^2} \overline{L_L} \varphi \sigma_{\mu\nu}
\tau_R B^{\mu\nu} ~,
\end{equation}
\begin{equation}
\label{eq:ow}
{\cal O}_W = \frac{1}{\Lambda^2} \overline{L_L} \vec{\tau}\varphi
\sigma_{\mu\nu} \tau_R \vec{W}^{\mu\nu}  ~.
\end{equation}
Here $L_L=(\nu_L,\tau_L)$ is the tau leptonic doublet, $\varphi$  is
 the Higgs doublet, $B^{\nu\nu}$ and $\vec{W}^{\mu\nu}$ are the 
U(1)$_Y$ and SU(2)$_L$ field strength tensors. 
Other dimension six operators like
\begin{equation}
\label{eq:oderiv}
\frac{1}{\Lambda^2} \overline{L_L} \sigma_{\mu\nu} \db L_L B^{\mu\nu} ~,
\end{equation}
reduce to the operator of \eq{eq:ob} after the use of the standard model 
equations of motion.
The effective Lagrangian is
\begin{equation}
\label{eq:leff}
{\cal L}_{eff} = \alpha_B {\cal O}_B + \alpha_W {\cal O}_W + \mathrm{h.c.}~,
\label{eq:interaccio}
\end{equation}
where we will take the couplings
$\alpha_B$ and $\alpha_W$ real. 

After spontaneous symmetry breaking, the Higgs gets a vacuum expectation
value  and, as usual,
the interactions in (\ref{eq:interaccio}) can be written in terms of the gauge 
boson mass eigenstates 
$A^\mu$, $Z^\mu$ and $W^\mu$.
Our Lagrangian, written in terms
of the mass eigenstates, is
\begin{eqnarray}
\nonumber
{\cal L}_{eff} &=& \hspace*{-2mm}
 \frac{a_\gamma e}{4 m_\tau}\overline{\tau} 
\sigma_{\mu\nu}\tau F^{\mu\nu} +
 \frac{a_Z e}{4 m_\tau}\overline{\tau} \sigma_{\mu\nu}
\tau
Z^{\mu\nu}\\ 
&+&\hspace*{-2mm}
\left( \frac{\kappa^W e}{4 \sqrt{2} m_\tau}  \overline{\nu_{\tau L}} 
\sigma_{\mu\nu}\tau_R W^{\mu\nu}_++\mathrm{h.c.}\right) 
\label{eq:leff_fin}
\end{eqnarray}
where one can recognize the usual couplings related to the magnetic moments.
$X_{\mu\nu}=\partial_\mu X_\nu-\partial_\nu X_\mu$ (for $X = F, Z,
W_+$)
 is 
the Abelian field strength tensor for the gauge bosons.
 We have not written the non-Abelian
couplings involving more than one gauge boson because they do not 
contribute to
magnetic moments at leading order. The above new physics magnetic moments
can be written as
\begin{eqnarray}
a_\gamma &\hspace*{-2mm}=& \hspace*{-2mm}(\alpha_B - \alpha_W) \frac{u \sqrt{2}
m_\tau}{\Lambda^2}~,\\
a_Z &\hspace*{-2mm}=& \hspace*{-2mm}- (\alpha_W c_W^2 + \alpha_B s_W^2) \frac{u \sqrt{2}
m_\tau}{s_W c_W \Lambda^2}~,\\
\kappa^W &\hspace*{-2mm}=&\hspace*{-2mm} \alpha_W \frac{u 2 \sqrt{2} m_\tau}{\Lambda^2}=
-2 s_W \left(c_W a_Z + s_W a_\gamma\right)
 \label{eq:epsilonw}
\end{eqnarray}
where $u/\sqrt{2}$ is the Higgs vacuum expectation value and $s_W=sin\theta_W$,
$c_W=cos\theta_W$.
The Lagrangian (\ref{eq:leff_fin}) gives additional contributions to
the anomalous standard model electromagnetic moment of the tau ${a_\gamma}^{SM}$. 
The same is true for the other  magnetic moments that have been
introduced for the  weak magnetic
moments: $a_Z$ for the $Z$-boson\cite{nos}, and $\kappa^W$
 for the $W$-boson\cite{rizzo}.
Sometimes it will be convenient to consider the following 
dimensionless couplings, proportional to the magnetic moments
\begin{eqnarray}
a_\gamma&=&2\, r_Z\, \eg ~,\\
a_Z&=&2\, \frac{1}{s_W\, c_W} r_Z\, \ez ~,\\
\kappa^W &=&  2 \sqrt{2} \, r_Z\, \epsilon_W ~.
\label{eq:moments}
\end{eqnarray}
where $r_Z = m_\tau/m_Z$.
We would like to stress the following point. In the effective 
Lagrangian approach, 
{\em exactly} the same couplings
that contribute to processes at high energies also
contribute to the (magnetic moment) form factors, $F^{\mathrm new}(q^2)$, 
at $q^2=0$. The
difference $F^{\mathrm new}(q^2)-F^{\mathrm new}(0)$ only comes from
higher dimension operators whose effect is suppressed by powers of 
$q^2/\Lambda^2$,  as
long as $q^2 \ll \Lambda$ as needed for the consistence of the effective
Lagrangian approach. This means that any measurement (bound) 
for the new physics contributions to magnetic moments, at an energy scale
 where some of the gauge bosons or other particles are off-shell,
can be directly added to the on-shell standard model prediction in
 order to get a value (limit) on the magnetic dipole moments.

The effects of the operators in (\ref{eq:leff_fin})
are suppressed at low energies. This means that  the most 
interesting bounds
will come from the highest precision experiments at the highest
available energies. At present they are LEP1 and SLD
($Z$ decay rates and polarization asymmetries),
LEP2 (cross sections and $W$ decays rates), CDF and D0 ($W$ 
decay rates). 
Consequently in the following section we will
study all those observables.

\section{OBSERVABLES}

In this section we will present the observables best suited
to set bounds  on the magnetic moments. 

\subsection{Universality in tau pair production 
 at LEP1 and SLD\label{sec:lep1}}

We begin with  tau pair production in $e^+ e^- \rightarrow \tau^+ \tau^-$
 from threshold to LEP2 energies. Therefore we will include 
both photon and Z-exchange with standard model vector and axial
 couplings to fermions,  plus additional magnetic moment couplings
 given by \eq{eq:leff_fin}. The expressions for the cross section 
can be found in \cite{npb,nos}.
It can be separated into 
three types of contributions:
i) the standard model tree level contribution , which
is the dominant one,
ii) a contribution which is proportional to the tau mass. 
This comes together
with an insertion of the magnetic moment operators 
(non-standard contribution)
 or with an insertion of another fermion mass
(standard model contributions) or both, two insertions of
magnetic moment operators and two mass insertions (non-standard
contributions), and 
iii) a contribution free of masses but with two insertions 
of the magnetic moment operators. 

This can be easily understood, since standard model couplings of gauge
bosons to fermions
conserve chirality, while mass terms and magnetic moment couplings change it,
therefore interference of magnetic moment contributions with standard
ones should be proportional to the fermion masses and only the square of
magnetic moments can be independent of fermion masses. In the limit of
zero tau mass only the contribution iii) is relevant, however there could
be some range of the parameters in which contribution ii) is higher
than iii). In fact for any finite value of the tau  mass it is obvious
that for large enough $\Lambda$ ii) will always dominate over iii). 
In order to be as general as possible we will include all three contributions.

Bounds on the couplings $a_\gamma$ and $a_Z$ can be obtained 
from LEP1-SLD universality tests by assuming that only the 
tau lepton has anomalous magnetic moments (muon and 
electron electromagnetic moments have been measured quite precisely
~\cite{caso}). In order to compare with experimental data  it 
is convenient to define the universality ratio:
\begin{eqnarray}
R_{\tau\mu}&=&\left.
\frac{\sigma(e^+e^-\rightarrow \tau^+\tau^-)}
                 {\sigma(e^+e^-\rightarrow
\mu^+\mu^-)}\right|_{s=m_Z^2}=
\frac{\Gamma_{\tau\bar{\tau}}}{\Gamma_{\mu\bar{\mu}}}\nonumber \\&=&
\frac{R_\mu}{R_\tau} \equiv R_{SM} + R_1 +R_2~.
\label{eq:ratiocross}
\end{eqnarray}
Here $R_\mu \equiv \Gamma_{\mathrm had}/\Gamma_{\mu\bar{\mu}}$ and
$R_\tau = \Gamma_{\mathrm had}/\Gamma_{\tau\bar{\tau}}$ are the quantities
directly measured  \cite{lepsld}, $R_{SM}$ is the standard 
model contribution (including lepton-mass corrections), and
$R_1$ and $R_2$ are the linear and quadratic terms, respectively, in
the tensor couplings. Notice that this quotient eliminates the hadronic
 part that,
although very well determined in experiments, has much higher theoretical
uncertainties than $R_{\tau\mu}$.
The theoretical  expression for $R_{\tau\mu}$ can 
be  computed from the cross sections, and it is
\begin{equation}
R_{SM} = \sqrt{1-4\, r_Z^2}\, 
 \left[1+2\, r_Z^2\, \frac{v^2-2a^2}{v^2+a^2}\right]~, 
\label{eq:ratio1}\end{equation}
\begin{equation}
R_1 = -12\, \sqrt{1-4\, r_Z^2}\, \,  r_Z\, 
\frac{v}{v^2+a^2}\, \ez~, 
\end{equation}
\begin{equation}
R_2 = 2\, \sqrt{1-4\, r_Z^2}\, \left[1+8\, r_Z^2\right]
\,\frac{1}{v^2+a^2}\, \ez^2\label{eq:rat2}
\end{equation}
with $a\equiv a_e=a_\tau=a_\mu$, $v\equiv v_e=v_\tau=v_\mu$.
 Notice that in the ratio \eq{eq:ratiocross} the 
electroweak radiative corrections cancel to a large extent 
and, therefore, we can use tree-level formulae. However, if needed,
the expressions in \eqs{eq:ratio1}{eq:rat2} can be improved by using
 effective couplings 
\cite{bernabeu}.
From the  very precise experimental LEP1 and SLD measurements \cite{lepsld}
 we obtain
\begin{equation}
R_{\tau\mu}=1.0011 \pm 0.0027~.
\label{eq:ratiocross1}
\end{equation}
Comparing the equation (\ref{eq:ratiocross1}) with 
(\ref{eq:ratiocross})
one gets 
\begin{equation}
0.0007 \leq 7.967\, \epsilon_Z^2+0.037\, \epsilon_Z\leq 0.0061
\label{eq:epsilonz1}
\end{equation}
that  leads to the following two bands for $\ez$: 
\begin{equation}\label{eq:epsilonz2}
-0.030\leq \epsilon_Z  \leq -0.012\quad \end{equation}
or
\begin{equation} 
0.007\leq \epsilon_Z \leq 0.025~,
\label{eq:ratiocross1limit}
\end{equation}

\subsection{Tau pair production  at {LEP2}\label{sec:lep2}}
At LEP2, the contributions coming
from the photon-exchange are dominant over those coming from
the $Z$-exchange, and as a result both magnetic moments,
 $a_\gamma$ and $a_Z$, will
enter into the constraints. 

Present experimental errors from $e^+e^-\rightarrow \tau^+\tau^-$ 
cross section
are much milder at LEP2  than at LEP1. A combination of the LEP2
data \cite{all} on this
cross section, for  $s'$  (the invariant mass of the pair 
of tau leptons) so that
$\sqrt{s'/s}>0.85$, is listed in table \ref{tabla1}. This 
combination of data has been only made for  the 183 GeV and 189 GeV 
data-sets as they have the highest
luminosity and center-of-mass energies. For comparison we also present
the standard model prediction for the cross section. In both, experimental
results and standard model predictions, initial-final state radiation photon
interference is subtracted.
\begin{table}
\caption{{\small Combined experimental data for the $\tau^+\, \tau^-$
cross section 
from ALEPH, DELPHI, L3, OPAL at LEP2 energies (in GeV). $\sqrt{s'/s}$ 
is the cut in the invariant mass of the tau pair.}}
\begin{tabular}{cccc}\hline
$\sqrt{s}$&$\sigma_{\tau\bar{\tau}}^{\rm
SM}$(pb)&$\sigma_{\tau\bar{\tau}}$(pb)&$\sqrt{s'/s}$\\\hline 
$182.7^{}$&3.45&$3.43\pm0.18$&0.85\\ 
188.6&3.21&$3.135\pm0.102$&0.85 \\ \hline\end{tabular}
\label{tabla1}
\end{table}
For LEP2 let us define the ratio $R_{\tau\bar{\tau}}$ as:
\begin{eqnarray}
R_{\tau\bar{\tau}}&\equiv& \frac{\sigma(e^+e^-\rightarrow \tau^+\tau^-)}{
\sigma(e^+e^-\rightarrow \tau^+\tau^-)_{SM}}=1\nonumber\\
&+& F_1^\gamma(s)\,  \eg  +
F_2^\gamma(s)\, \eg^2+F_1^Z(s)\, \ez\nonumber\\ 
&+&F_2^Z(s)\,  \ez^2 +F^{\gamma Z}(s)\, \ez \eg
\label{eq:ratiocross2}
\end{eqnarray}
For the range of energies used by LEP2 experiments, the
coefficients $F_i(s)$, obtained from the cross sections, are given in 
table \ref{tabla2}. Direct comparison of \eq{eq:ratiocross2} to 
experimental data will provide bounds on the anomalous couplings. We have
checked that, even though coefficients in table \ref{tabla2} are obtained
with no initial state radiation, its inclusion only changes 
the coefficients by about a 10\% and  
this does not affect significantly the obtained bounds.

\begin{table}
\caption{{\small Coefficients of the anomalous contributions to 
$R_{\tau\bar{\tau}}$, for 
the different center of mass measured energies (in GeV) at LEP2.}}
\begin{tabular}{cccccc}
\hline
$\sqrt{s}$&$F^\gamma_1$&$F^\gamma_2$&$F^Z_1$&$F^Z_2$&
$F^{\gamma Z}$\\
\hline
130&0.079&0.682&0.028&5.258&0.286\\ 
136&0.083&0.784&0.026&5.152&0.304\\ 
161&0.092&1.221&0.022&5.272&0.384\\ 
172&0.094&1.427&0.021&5.497&0.424\\ 
183&0.096&1.642&0.020&5.789&0.467\\ 
189&0.096&1.765&0.019&5.971&0.491\\ \hline
\end{tabular}
\label{tabla2}
\end{table}

All data are used 
independently in the global fit discussed in the final section.
Just as an example of how well the new couplings
can be bound from LEP2 let us find the limits on $\eg$ obtained
by using only the data at $189$~GeV.  The
experimental value for the ratio $R_{\tau\bar{\tau}}$ is in this case:
\begin{equation}
\left.R_{\tau\bar{\tau}}\right|_{\rm exp}=0.978\pm0.032~,
\end{equation}
to be compared with the theoretical prediction (assuming that $\ez$ is
well bounded from LEP1-SLD, as it is)
\begin{equation}
\left.R_{\tau\bar{\tau}}\right|_{\rm th}=1.00+1.765\eg^2+0.096\eg~.
\end{equation}
From the two previous equations we find:
\begin{equation}
-0.10< \eg < 0.05~,
\label{eq:ratiocross2limit}                 
\end{equation}
This $1\sigma$  bound  is comparable to the one obtained in
the global fit given in the final section where 
all available data have been included.

\subsection{Tau lepton transverse polarization\label{sec:pol}}
 Chirality flipping observables vanish
for massless taus and depend linearly on magnetic
moments. On the other hand
they will not get contributions from physics conserving chirality.
In that sense they are truly magnetic moment observables. 
At LEP1, with the $\tau$ direction fully reconstructed
in the semi-leptonic decays
it has been shown\cite{nos}  that  one can measure
the anomalous weak magnetic moment
$a_Z$ by
measuring  an azimuthal asymmetry on the 
tau and tau decay products angles. This asymmetry selects the leading term
in the weak magnetic moment in the tau pair production cross section,
 that appears in the P-odd, 
chirality flipping  transverse tau polarization.
The expression  one finds 
for the proposed asymmetry is:
\begin{eqnarray}
&&\spa A_{cc}^\mp= \mp \alpha_h 
\frac{1}{2}\frac{a}{v^2+a^2}
\left[  
-v\, r_Z +\epsilon_Z 
\right]~,\label{l3}
\end{eqnarray}
where $\alpha_h=(m_\tau^2-2m_h^2)/((m_\tau^2+2m_h^2)$, is the polarization
analyzer for each hadron channel ($h=\pi,\rho$), $a\equiv a_e=a_\tau$,
and $v\equiv v_e=v_\tau$. 
In addition, the SM contribution to this observable is doubly suppressed
with respect to the non-standard one: by the
fermion-boson  vector coupling $v$  and by the $r_Z$  factor.

Within $1\sigma$, the
LEP1  measurement of this asymmetry and the SLD
determination of the transverse tau polarization\cite{l3}, 
translate into the following values for the $\epsilon_Z$ coupling
\begin{equation}
\epsilon_Z= \left\{\begin{array}{l}
(0.0\pm1.7\pm2.4)\times 10^{-2}\;(\mathrm{LEP1})~,\cr
(0.28\pm1.07\pm0.81)\times 10^{-2}\; (\mathrm{SLD})
\end{array}\right.
\end{equation}

Combining these results one gets the bound:

\begin{equation}
\epsilon_Z= 0.002\pm 0.012~. \label{eq:ass1}
\end{equation}

Note that even though the transverse tau polarization has been 
measured at LEP1-SLD
with a precision one order of magnitude worse than the universality test
$R_{\tau \mu}$  (2-4\% typically for the asymmetry, and 0.5\% for the
tau-muon cross section ratio), the obtained bound \eq{eq:ass1} is
as good as the  one coming from universality \eq{eq:ratiocross1limit}.
This is  so because the asymmetry depends linearly on the couplings.

At present there does not exists a similar measurement at LEP2. This would
 allow  to 
disentangle  the $\gamma$ components from the $Z$ components of the 
magnetic moments.

\subsection{Lepton universality in W decays.\label{sec:wdec}} 

 The Lagrangian \eq{eq:leff_fin} shows that the same couplings 
that give rise to
electromagnetic and $Z$-boson magnetic moments, also contribute  to the
couplings of the $W$ gauge bosons to tau leptons. The couplings appear in a
different combination than that in the photon or $Z$ couplings, so their
study gives  additional independent information on magnetic moment
couplings. As was noticed in Ref. \cite{rizzo} the best place to look for
effects of the $\epsilon_W$ coupling is in the $W$ decay widths.

Using our effective Lagrangian we can easily compute the ratio of the decay
width of the W-gauge boson in tau-leptons (with magnetic moments) to
the decay width of the W to electrons (without magnetic moments).
\begin{eqnarray}
R^W_{\tau e} &\equiv& \frac{\Gamma(W\rightarrow \tau\nu)}
{\Gamma(W\rightarrow e\nu)}
=(1-r_W^2)^3\times\nonumber\\&&\hspace*{-1.6cm}\left[ 1+\frac{r_W^2}{2}+
3\sqrt{2}\; r_W\ c_W\, \epsilon_W 
+(1+2\, r_W^2)\epsilon_W^2\right]
\label{eq:ratiow}
\end{eqnarray}
where $r_W = m_\tau/m_W$, and $\epsilon_W$ can be rewritten 
in terms of $\epsilon_\gamma$ and $\epsilon_Z$ as in \eq{eq:epsilonw}. 
Note that $R^W_{\tau e}$, like the cross
sections studied in subsection (1), is a chirality even observable.
The decay of the $W$ into leptons has been measured to a rather
good precision at LEP2, UA1, UA2, CDF and D0. There, results are presented 
in the form  of universality tests on the couplings \cite{lepsld,pich} 
that we rewrite  as a measurement on the
ratio $R^W_{\tau e}$ defined above
\begin{equation}
R^W_{\tau e} = 1.002 \pm 0.030~,
\label{eq:ratiowexp}
\end{equation}
Then, from \eqs{eq:ratiowexp}{eq:ratiow},  
we obtain the following limit on the $W$-boson magnetic moments.
\begin{equation}
-0.23\leq \epsilon_W\leq 0.15~.
\label{eq:wlimit}
\end{equation}

\section{COMBINED ANALYSIS AND RESULTS}

We have performed a global fit, as a function of the two independent
couplings $\epsilon_\gamma$ and $\epsilon_Z$, to the  studied 
observables : 
\begin{itemize}
\item Lepton universality
$R_{\tau\mu}=\frac{\sigma(e^+e^-\rightarrow \tau^+\tau^-)}
                 {\sigma(e^+e^-\rightarrow
\mu^+\mu^-)}$ at LEP1 and SLD; 
\item  the ratio of cross sections
$R_{\tau\bar{\tau}}\equiv \frac{\sigma(e^+e^-\rightarrow \tau^+\tau^-)}{
\sigma(e^+e^-\rightarrow \tau^+\tau^-)_{SM}}$,
for the two highest energies measured at LEP2;
\item the transverse tau polarization and  
polarization asymmetry
$A_{cc}^\mp$  measured at SLD and LEP1 ;
\item  lepton universality of W decays
$R^W_{\tau e} \equiv \frac{\Gamma(W\rightarrow \tau\nu)}
{\Gamma(W\rightarrow e\nu)}$
 measured at LEP2 and $p\bar{p}$ colliders. 
\end{itemize}

\begin{figure}[hbtp]
\begin{center}
\includegraphics[scale=0.28]{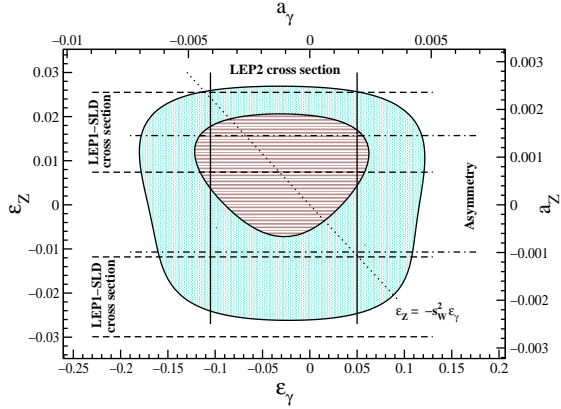}
\end{center}
\caption{{\small Global fit including all constraints;
95\% CL and 68\% CL contours are shown. The bands between straight lines 
show the allowed  1$\sigma$ regions coming from the different experiments: 
solid (LEP2-189 GeV), dashed (LEP1-SLD cross section), dot-dashed (asymmetry).
We also have plotted the line
$\epsilon_Z = - s^2_W \epsilon_\gamma$ (dotted line). 
This relationship appears when
only the operator ${\cal O}_B$ contributes.}
\label{fig:gfit}}
\end{figure}

In fig.~\ref{fig:gfit} we present, in the plane $a_\gamma$--$a_Z$ (or
$\eg$--$\ez$)
the allowed region of parameters at 1$\sigma$ and 2$\sigma$.
For comparison we also present (at 1$\sigma$) the relevant limits set independently
by the different observables, as discussed in the text.
By projecting onto the axes one can read off the 1$\sigma$ and 2$\sigma$
limits on the different  non-standard contributions to 
the anomalous electromagnetic and weak
magnetic moments $a_\gamma$, $a_Z$ 
\begin{eqnarray}
&&(1\sigma)\rightarrow \left\{
\begin{array}{ll}
 -0.005< a_\gamma <0.002~, \\ 
 -0.0007 < a_Z < 0.0019~,  
\end{array}
\right.\\
&&(2\sigma)\rightarrow \left\{
\begin{array}{ll}
 -0.007 < a_\gamma < 0.005~, \\ 
 -0.0024 < a_Z <0.0025~.  
\end{array}
\right.\label{eq:final2sigma}
\end{eqnarray}

Using the relationship among $\epsilon_\gamma$, $\epsilon_Z$,
$\alpha_B$ and $\alpha_W$ at a given value of the scale of new physics,
one can easily obtain bounds on $\alpha_B$ and $\alpha_W$. Alternatively,
by assuming that $\alpha_B/4\pi$ or $\alpha_W/4\pi$ 
are order unity one finds a
 bound on the scale of new physics: $\Lambda >9$~TeV.

We note that, for the first time,
 the bound for $a_\gamma$ is of the order of magnitude of
the prediction computed long ago by Schwinger $a_\gamma^{QED} \sim 0.0012$.

Bounds on the anomalous electromagnetic moment
for the $\tau$ have been considered and measured from the radiative 
decay $Z\rightarrow \tau^+\ \tau^- \ \gamma$ at LEP1
\cite{mendez,swain}. 
There, only the anomalous coupling $a_\gamma$ is taken into account,
 while the contributions coming from the tau $Z$-magnetic coupling $a_Z$ or 
the effective 4-particle vertex are neglected. The interpretation
of off-shell form factors is problematic since they can hardly be isolated
from the other contributions and gauge invariance can be a problem. In
the effective Lagrangian approach all those problems are solved because
form factors are directly related to couplings in the effective Lagrangian,
which is gauge invariant, and as discussed above,
the difference $F^{\mathrm new}(q^2)-F^{\mathrm new}(0)$ only comes from
higher dimension operators whose effect is suppressed by $q^2/\Lambda^2$. 
Using this approach,  the analysis of the L3  and OPAL  collaborations lead to the PDG
$95\%$ CL limit\cite{caso}:
\begin{equation} -0.052<a_\gamma<0.058 \end{equation} 
As can be seen from \eq{eq:final2sigma} our result, coming mainly from
LEP2, is about one order of magnitude better than the ones obtained 
from the radiative $Z$-decay.

The standard model prediction for the weak magnetic moment $a_Z$, which
was 
computed in \cite{nos}, is far below  the present bound.

The above bounds are completely model independent and no assumption has 
been made on the relative size of couplings $\alpha_B$ and $\alpha_W$ in the
effective Lagrangian (\ref{eq:leff}). For the sake of comparison with
published data \cite{masso} we present now the limits that can
be found by considering separately  only operator ${\cal O}_B$ or only
operator ${\cal O}_W$ in the Lagrangian (\ref{eq:leff}). Consider that only
${\cal O}_B$ is present, as in Ref.~\cite{masso}, is equivalent 
to impose the relation $\ez=-s_W^2\eg$ or equivalently $c_W\,a_Z = - s_W\, a_\gamma$.
 Thus, from fig.~\ref{fig:gfit}, it is
straightforward to obtain that the bounds on the anomalous magnetic 
moment (at $2\sigma$) are reduced to $-0.004 < a_\gamma < 0.003$, while little 
change is found on the weak-magnetic moment $-0.0019 < a_Z < 0.0024$.


Universality tests in $W$ decays do not provide any interesting constraint
on  $a_Z$ and $a_\gamma$. In fact, the  straight lines coming from the direct 
1$\sigma$ bound from universality tests in $W$ decays lie well outside the figure.
However, because of the relationship
(\ref{eq:epsilonw}), the LEP1-SLD and LEP2 constraints on $\ez$ and $\eg$ can
be translated into constraints on $\epsilon_W$ (or
$\kappa^W$ defined in \eq{eq:moments}),
the weak magnetic moment couplings of the $W$-gauge-boson to taus and
neutrinos. 
One  obtains the 95\% CL limits
\begin{equation}
-0.003 < \kappa^W < 0.004~.\label{eq:kweak}
\end{equation}

We have shown that the use of all available data at the highest
available energies (LEP1, SLD, LEP2, D0, CDF) leads to  strong 
constraints in all the magnetic moments ($\gamma , W, Z$) of the tau lepton
without making any assumption about naturalness or fine tuning. The obtained
bounds (\eq{eq:final2sigma} and \eq{eq:kweak}) are, to our knowledge,   
the best bounds that one can find in published data.


\end{document}